\documentclass[aps,prd,a4paper,amsmath,amssymb,nofootinbib,showpacs,showkeys,
preprintnumbers,notitlepage,superscriptaddress,twocolumn]{revtex4-1}

\usepackage{bm}
\usepackage{graphicx}

\begin{document}

\title{Polyakov Loop and Gluon Quasiparticles in Yang-Mills Thermodynamics}

\author{M. Ruggieri}\email{Corresponding author: marco.ruggieri@lns.infn.it}
\affiliation{Department of Physics and Astronomy, University of Catania, Via S. Sofia 64, I-95125 Catania}
\author{P. Alba}
\affiliation{INFN-Laboratori Nazionali del Sud, Via S. Sofia 62, I-95123 Catania, Italy}
\author{P. Castorina}%\email{}
\affiliation{Department of Physics and Astronomy, University of Catania, Via S. Sofia 64, I-95125 Catania}
\affiliation{INFN, Sezione di Catania, I-95123 Catania, Via Santa Sofia 64, Italy.}
\author{S. Plumari}%\email{}
\affiliation{Department of Physics and Astronomy, University of Catania, Via S. Sofia 64, I-95125 Catania}
\affiliation{INFN-Laboratori Nazionali del Sud, Via S. Sofia 62, I-95123 Catania, Italy}
\author{C. Ratti}%\email{ruggieri@yukawa.kyoto-u.ac.jp}
\affiliation{Dipartimento di Fisica, Universita` degli Studi
di Torino e INFN, Sezione di Torino, via Giuria 1, I-10125 Torino (Italy)}
%\affiliation{Department of Physics and Astronomy, University of Catania, Via S. Sofia 64, I-95125 Catania}
\author{V. Greco}%\email{greco@lns.infn.it}
\affiliation{Department of Physics and Astronomy, University of Catania, Via S. Sofia 64, I-95125 Catania}
\affiliation{INFN-Laboratori Nazionali del Sud, Via S. Sofia 62, I-95123 Catania, Italy}

%************************************   Abstract   ********************************************%

\begin{abstract}
We study the interpretation of Lattice data about the thermodynamics
of the deconfinement phase of $SU(3)$ Yang-Mills theory, in terms of gluon quasiparticles
propagating in a background of a Polyakov loop. A potential for the Polyakov loop,
inspired by the strong coupling expansion of the QCD action, is introduced; 
the Polyakov loop is coupled to tranverse gluon quasiparticles by means of a gas-like effective potential.
This study is useful to identify the effective degrees of freedom propagating in the gluon medium above the
critical temperature.
A main general finding is that a dominant part of the phase transition dynamics
is accounted for by the Polyakov loop dynamics, hence the thermodynamics can be described without the need for diverging
or exponentially increasing
quasiparticle masses as $T \rightarrow T_c$, at variance respect to standard quasiparticle models.
\end{abstract}

%\preprint{Catania-12-XY}

\maketitle

\section{Introduction}
Since the seminal paper by Yang and Mills~\cite{Yang:1954ek}, the study of
nonabelian gauge theories has attracted a lot of interest in several sectors
of physics. Focusing on the gauge theory of strong interactions (QCD), 
thanks to improvement of computer facilities which allow to perform
huge simulations of Yang-Mills theories, as well as to the 
possibility to create in laboratories extremely hot environments
by means of heavy ion collisions, the interest in understanding the
thermodynamical properties of strong nonabelian mediums has noticeably increased.

Lattice simulations of $SU(3)$ Yang-Mills 
theories have been performed, see for example Refs.~\cite{Datta:2010sq,Boyd:1996bx,Boyd:1995zg,Borsanyi:2011zm}.
All of these studies agree about the onset of the deconfinenemt phase transition
at $T= T_c \approx 270$ MeV. Below $T_c$, the thermodynamics of the 
Yang-Mills theory is dominated by the lowest 
lying glueballs. For what concerns the high temperature phase, the picture
is not so clear. As a matter of fact,  one of the most intriguing aspects is the nonperturbative
nature of the gluon medium above the critical temperature \cite{zwanziger2}. For example,
in~\cite{Borsanyi:2011zm} it is pointed out that the perturbative regime
is realized for temperatures $T \gg 100 T_c$ (more concretely, at $T=100 T_c$ 
both pressure and energy density are below the Stefan-Boltzmann limit, which
corresponds to the ideal gas case, by about $10\%$). This makes the identification
of the correct degrees of freedom of the gluon plasma, in proximity
as well as well beyond the critical temperature, a very complicated task.
Resummation schemes have been proposed, based for example on the Hard Thermal Loop (HTL) approach~\cite{HTL1,HTL3,HTL4,HTL7}. 
More recently within a NNLO in the HTL perturbation theory \cite{Andersen_HTLpt} it has been shown that an agreement with the lattice 
results can be obtained also for the trace anomaly down to  $T \simeq 2\,T_c$ but with large
uncertainties. At these high temperatures, the HTL 
approach motivates and justifies a picture of weakly interacting quasi-particles, as 
determined by the HTL propagators.

Within this theoretical framework, the quasiparticle approach to the thermodynamics of QCD has attracted a discreet 
interest recently, see for example Refs.~\cite{Gorenstein:1995vm,PKPS96, Levai:1997yx,bluhm-qpm,
Meisinger:2003id,Castorina:2011ra,Giacosa:2010vz,Peshier:2005pp,redlich,
Plumari:2011mk,Filinov:2012pt,Castorina:2007qv,Castorina:2011ja,Brau:2009mp,Cao:2012qa,Bannur:2006hp} and
references therein. In such an approach, one identifies the degrees of 
freedom of the deconfinement phase with transverse gluons;
the strong interaction in such a non perturbative regime
is taken into account through an effective temperature-dependent mass for the gluons themselves.
This theoretical approach is similar to the one introduced by
Landau~\cite{landau9} to describe strongly correlated
systems. Generally speaking, it is not possible to compute the
self-energy of gluons exactly in the range of temperature
of interest; as an obvious consequence, the location of the poles
of the propagator in the complex momentum plane, hence (roughly
speaking) the gluon mass, is an unknown function. For this reason,
one usually assumes an analytic dependence of the gluon mass on the
temperature, leaving few free parameters which are then fixed
by fitting the thermodynamical data of Lattice simulations. 

The advantage of such an approach is, at least, twofolds. 
Firstly, it is of a theoretical interest in itself to understand
which are the effective degrees of freedom of the Yang-Mills 
theory at finite temperature and in particular the evolution of
non perturbative effects with temperature.
Furthermore, once a microscopic description of the plasma is identified, 
it is also possible to include the implied dynamics into a transport theory capable to directly 
simulate the expanding fireball produced in heavy ion collisions
computing the collective properties, as well as the chemical 
composition of the fireball as a function of time~\cite{Scardina:2012hy}. 

In this article we extend the quasiparticle picture of the finite
temperature gluon medium, by adding an interaction of gluons 
with a background Polyakov loop. It is well known that the latter 
is an exact order parameter for the confinement-deconfinement
transition in the pure gauge theory~\cite{Polyakov:1978vu,Susskind:1979up,Svetitsky:1982gs,Svetitsky:1985ye}. 
Therefore, having access to the Polyakov
loop immediately leads us to the possibility to study the effect of the
phase transition on the dynamical properties of gluons, mainly
on their mass.
The standard quasiparticle approach, instead, tries to account for all the dynamics 
only by mean of temperature dependent masses \cite{PKPS96,Levai:1997yx, bluhm-qpm,redlich,Castorina:2011ra,
Castorina:2011ja, Plumari:2011mk}. This always leads to the necessity of diverging
or steadily increasing masses
as $T\rightarrow T_c$ not only of the case of pure gauge theory, but also for the QCD
case which includes quarks.
We show that combining a T-dependent quasiparticle mass, $m_g(T)$, with the Polyakov
loop dynamics results in a quite different behavior of $m_g(T)$ as $T\rightarrow T_c$.

The combined approach of the quasiparticle picture with the Polyakov loop
thermodynamics has been poorly investigated. 
To our knowledge, there is only a first preliminary study about the role of the Polyakov
loop in the context of the quasiparticle description of the gluon
plasma ~\cite{Meisinger:2003id}, which has some overlap with our study.
In the present article, we take a simple matrix model action as 
the effective potential for the Polyakov loop. 
Then, we use the Weiss mean field procedure to
compute averages. Such a procedure has been used in the study 
of the QCD phase diagram by means of the Polyakov extended Nambu-Jona Lasinio 
model~\cite{Fukushima:2003fw,Ratti:2005jh}, see~\cite{Abuki:2009dt,Zhang:2010kn}. 
In the next Section we describe the technical details, it suffices to say here that 
the main difference between this procedure and the one commonly used in the literature
is that, in the former case, averages are computed in terms
of integrals over the gauge group with the appropriate
invariant gauge measure having the pure gauge mean field action as
the integration weight.  We can anticipate that in such a way we are not only able to reproduce
the large trace anomaly, $T_\mu^\mu=\epsilon - 3 P$,  with a quite good accuracy,
but also identify the different roles of quasiparticles masses and Polyakov loop
in Yang-Mills thermodynamics.

The paper is organized as follows.
In Section II, we describe the matrix model in the Weiss mean field potential and its
coupling to quasiparticle masses, showing the explicit derivation for $SU(3)$ gauge group.
In Section III, we compare the results of our approach to the lQCD thermodynamics
discussing in particular the specific contributions to the pressure and to the interaction measure
of the Polyakov mean filed and of the quasiparticle masses .
In Section IV we draw first conclusions of our work along with future possible investigations and developments
of the present approach.

\section{The model}
The system we consider in this article consists of a gas of gluon quasiparticles,
propagating in a background of Polyakov loop. The free energy of the model
is expressed as a linear combination of two contributions:
the first one describes the thermodynamics of the Polyakov loop; the second one,
on the other hand, is the contribution of gluon quasiparticles coupled
to the Polyakov loop. We specify the two terms of the effective potential
below. 

\subsection{The matrix model in the Weiss mean field potential}
In this Section we follow closely the notations of~\cite{Abuki:2009dt,Zhang:2010kn}. 
The starting point is the action of a matrix model~\cite{Gupta:2007ax}, 
\begin{equation}
S_{PM}[L] = -N_c^2 e^{-a/T}\bm\sum_{\bm x,\bm y} \ell_F(\bm x)\ell_F^*(\bm x + \bm y)~,
\label{eq:sc1}
\end{equation}
where $\bm x,\bm y$ denote the lattice site and its nearest neighbors respectively. Moreover, 
\begin{equation}
\ell_F(\bm x) = \frac{1}{N_c} \text{Tr}L_F(\bm x)~,
\label{eq:PL}
\end{equation}
corresponds to the traced Polyakov loop in the fundamental representation. The Polyakov line
in the representation ${\cal R}$ of the gauge group is given by
\begin{equation}
L_{\cal R}(\bm x) = {\cal P}\exp\left[i\int_0^{1/T}d\tau~A_4^a(\bm x,\tau)T_{a,{\cal R}}\right]~,
\label{eq:RRR}
\end{equation}
where $T_{a,{\cal R}}$ are the gauge group generators in the representation ${\cal R}$.

The model specified by Eq.~\eqref{eq:sc1} corresponds to a matrix model, the untraced Polyakov loops 
$L(\bm x)$ corresponding to the dynamical degrees of freedom. The partition function can be written as
\begin{equation}
{\cal Z} = \exp(-\Omega_{PM}/T) = \int\bm\prod_{\bm x} dL(\bm x)\exp(-S_{PM}[L])~.
\label{eq:PF}
\end{equation}
Here, $dL$ is the group invariant measure on the $SU(N_c)$ gauge group. It can be written as
\begin{equation}
dL = \prod_{i=1}^{N_c -1} d\theta_i\prod_{i<j}^{N_c}\left|e^{i\theta_i} - e^{i\theta_j}\right|^2~
\delta(\theta_1 + \theta_2 + \dots \theta_{N_c}),
\label{eq:HM}
\end{equation}
and the integration over the variables $\theta_i$ is performed over the range $[0,2\pi]$.

Adopting the Weiss mean field procedure, the action in Eq.~\eqref{eq:sc1} can be written as
\begin{equation}
S_{mf} = -N_c\bm\sum_{\bm x}\left[\alpha\Re\ell_F + i\beta\Im\ell_F\right]~,
\label{eq:MFact}
\end{equation}
where $\alpha,\beta$ denote the mean fields of the Polyakov loop and of its conjugate respectively.
In terms of $S_{mf}$, the average of an operator ${\cal O}$, namely $\langle{\cal O}\rangle$, is defined as
\begin{equation}
\langle{\cal O}\rangle = \frac{\int\bm\prod_{\bm x} dL(\bm x){\cal O}[L]e^{-S_{mf}}}
{\int\bm\prod_{\bm x} dL(\bm x) e^{-S_{mf}}}~.
\label{eq:MF}
\end{equation}

The thermodynamic potential $\Omega_{PM}$ can be written by adding and subtracting the mean field action $S_{mf}$
in the exponent of Eq.~\eqref{eq:PF}. We have
\begin{equation}
\frac{\Omega_{PM}}{T} = -\log\langle e^{-(S_{PM} - S_{mf})}\rangle
- \log\int\bm\prod_{\bm x} dL(\bm x) e^{-S_{mf}}~,
\label{eq:21}
\end{equation}
where the average has to be performed as in Eq.~\eqref{eq:MF}.

\begin{widetext}
Equation~\eqref{eq:21} is exact. In the mean field approximation, we replace $\langle e^{-(S_{PM} - S_{mf})}\rangle$
with $ e^{-\langle S_{PM} - S_{mf}\rangle}$. In this way, the thermodynamic potential depends explicitely on the
mean fields $\alpha,\beta$, and their values are determined selfconsistently by the stationarity condition.
 In the mean field approximation the thermodynamic potential can thus be written as~\cite{Abuki:2009dt,Zhang:2010kn}
\begin{eqnarray}
\frac{\Omega_{PM} a_s^3}{T V} &=& -2(d-1)N_c^2 e^{-a/T}\langle\ell_F\rangle\langle\ell_F^*\rangle \nonumber\\
&&+\frac{N_c}{2}\left[(\alpha+\beta)\langle\ell_F\rangle + (\alpha-\beta)\langle\ell_F^*\rangle\right] \nonumber \\
&&-\log\int dL~e^{N_c\left[\alpha\Re\ell_F + i\beta\Im\ell_F\right]}~.
\label{eq:MFW}
\end{eqnarray}
In the previous equation, $d$ corresponds to the spacetime dimension, so that $2(d-1)$ is the coordination number
of the cubic lattice, that is the number of nearest neighbors of the site $\bm x$. Furthermore, $a_s$ is the
lattice spacing, and $V$ is the volume of the lattice; as a consequence, $V/a_s^3$ is the number of lattice sites. 

%\begin{widetext}
The thermal average of the Polyakov loop and of its conjugate are given by
\begin{eqnarray}
\langle\ell_F\rangle &=& \frac{\int dL~e^{N_c\alpha\Re\ell_F}
\left[\cos(N_c\beta\Im\ell_F)\Re\ell_F - \sin(N_c\beta\Im\ell_F)\Im\ell_F\right]}
{\int dL~e^{N_c\alpha\Re\ell_F}\cos(N_c\beta\Im\ell_F)}~,\\
\langle\ell_F^*\rangle &=& \frac{\int dL~e^{N_c\alpha\Re\ell_F}
\left[\cos(N_c\beta\Im\ell_F)\Re\ell_F + \sin(N_c\beta\Im\ell_F)\Im\ell_F\right]}
{\int dL~e^{N_c\alpha\Re\ell_F}\cos(N_c\beta\Im\ell_F)}~;
\label{eq:bla}
\end{eqnarray}
From the above equations it is then clear that $\langle\ell_F\rangle$ and $\langle\ell_F^*\rangle$
are different when $\beta\neq0$. This happens for example in theories with fermions at finite
baryon chemical potential.

In the following we focus on the case of the pure gauge theory, in which $\langle\ell_F\rangle = \langle\ell_F^*\rangle$. 
We have thus $\beta=0$, and Eq.~\eqref{eq:MFW} can be written as
\begin{eqnarray}
\Omega_{PM} &=& b T\left(-2(d-1)N_c^2 e^{-a/T}\langle\ell_F\rangle^2
+N_c\left[\alpha \langle\ell_F\rangle \right] 
-\log\int dL~e^{N_c\left[\alpha \Re \ell_F \right]}\right)~,
\label{eq:MFWreal}
\end{eqnarray}
where we have used the shorthand notation $b= V/a_s^3$. In the following, $a$ and $b$ in the above equation
are treated as free parameters, which will be fixed, together with the remaining two parameters
of the model which will be described in the next Section, by a minimization procedure of the
mean quadratic deviation between the theoretical computations and the Lattice data or pressure,
energy density and interaction measure.
For the sake of future reference, we call the model described by the above thermodynamic
potential as the pure matrix model (PMM).

\end{widetext}

\subsection{The quasiparticle gas contribution}
The gluon quasiparticle contribution to the thermodynamic potential reads~\cite{Meisinger:2003id}
\begin{equation}
\Omega_{qp} = 2T\int\frac{d^3k}{(2\pi)^3}\text{Tr}_A\log\left(1-L_A~e^{-E(k)/T}\right)~.
\label{eq:gqp}
\end{equation}
In the above equation, $L_A$ corresponds to the Polyakov line in the adjoint representation,
as defined in Eq.~\eqref{eq:RRR}, and the trace is taken over the indices of the adjoint representation
of the gauge group. Moreover, the quasiparticle energy is given by $E(k) = \sqrt{k^2 + M^2}$,
where $M$ is supposed to arise from non-perturbative medium effects. 
In the case of very high temperature, where the gauge theory is in the perturbative regime and thus $M = 0$,
it can be proved that Eq.~\eqref{eq:gqp} is the effective potential for the adjoint 
Polyakov loop~\cite{Weiss:1980rj}. At lower temperatures, where non-perturbative effects are important,
Eq.~\eqref{eq:gqp} must be postulated as a starting point for a phenomenological description of the 
thermodynamics of the gluon plasma. In this article, we make use of a temperature depentent mass whose analytic form
is given by~\cite{Peshier:2005pp}
\begin{equation}
M^2 = \frac{8\pi^2 T^2}{11\log\left[\lambda(T-w)\right]^2}~.
\label{eq:mass}
\end{equation}
The parameters $w,\lambda$ are determined by requiring that, for pressure and energy density, the mean quadratic deviation 
between our theoretical computation and the Lattice data of~\cite{Boyd:1996bx} is a minimum.
In particular, we find $w=240$ MeV and $\lambda = 0.25$, see Section III.
The temperature dependent quasi-particle mass in Eq.~\eqref{eq:mass} has several motivations:
firstly, in the very large temperature regime in which QCD is perturbative, one has $M = g(T) T$;
moreover, for $T$ close to $T_c$ there are remnants of the confining mechanism and,
a priori, there is no reason to reduce these effects to the Polyakov loop dynamics;
finally, for $T \rightarrow T_c^-$, the glueball mass decreases~\cite{Brau:2009mp}, 
$M_{glueball} \approx 1$ GeV at $T=T_c$ , and in a valence gluon model
one expects an effective mass of the quasiparticle $M \approx 500$ MeV at the melting temperature.

In order to have a combined description of the gluon plasma around the critical temperature,
in terms of the Polyakov loop (which acts as a background field) and of gluons quasiparticles
(which propagate in the Polyakov loop background), we add Eq.~\eqref{eq:gqp} to
the strong coupling inspired potential in Eq.~\eqref{eq:MFWreal}. In order to do this,
and to be consistent at the same time with the Weiss mean field procedure outlined above,
we follow the lines of~\cite{Abuki:2009dt} and write the thermodynamic potential as
\begin{equation}
\Omega = \Omega_{PM} + \langle\Omega_{qp}\rangle~,
\label{eq:FULL1}
\end{equation} 
where the average must be understood as defined in Eq.~\eqref{eq:MF}. In taking the average,
we use the same approximation of~\cite{Abuki:2009dt} and write
\begin{equation}
\left\langle\log\det\left(1-L_A~e^{-E/T}\right)\right\rangle = 
\log\left\langle\det\left(1-L_A~e^{-E/T}\right)\right\rangle~.
\label{eq:UUU}
\end{equation}
This approximation is also discussed in the Appendix B of~\cite{Zhang:2010kn}, where it is proved
that taking the average of the log (that is, $\langle\Omega_{qp}\rangle$) instead of the argument of the log, 
would lead to unphysical artifacts in the case the particles are taken in representations
different from the fundamental one. We will discuss this aspect in more detail in the next Section.

\subsection{The $SU(3)$ case}
In the case of the color group $SU(3)$ the Polyakov line can be parametrized in terms
of two angles, $\theta_3$ and $\theta_8$, namely $L_F = \text{diag}(e^{i\theta_3},e^{i\theta_8},e^{-i(\theta_3 + \theta_8)})$;
the invariant Haar measure reads:
\begin{equation}
dL = \frac{d\theta_3 d\theta_8}{6\pi^2}\left[\sin(\theta_3 - \theta_8) - \sin(\theta_3 - \theta_9) 
+ \sin(\theta_8 - \theta_9)\right]^2~,
\label{eq:HaarSU3}
\end{equation}
with $\theta_9 = -(\theta_3 + \theta_8)$. We define
\begin{eqnarray}
F(\alpha) &=& \int dL~e^{\alpha N_c \Re\ell_F}=\int dL e^{\alpha \Re f(\theta_3,\theta_8)}
\end{eqnarray}
where
\begin{eqnarray}
\ell_F &=& \frac{1}{N_c}\left(e^{i\theta_3} + e^{i\theta_8} 
+ e^{-i(\theta_3 + \theta_8)}\right) \equiv\frac{1}{N_c}f(\theta_3,\theta_8)~.\nonumber\\
&&
\end{eqnarray}
Moreover we have
\begin{equation}
\langle\ell_F\rangle =
\frac{1}{N_c}\frac{\int dL e^{\alpha \Re f(\theta_1,\theta_2)} f(\theta_1,\theta_2)}
{F(\alpha)}~.
\end{equation}
The thermodynamic potential for the pure matrix model in this case then reads
\begin{eqnarray}
\Omega_{PM} &=& bT\left(-6 N_c^2 e^{-a/T}\langle\ell_F\rangle^2 
+N_c~\alpha \langle\ell_F\rangle 
-\log F(\alpha)\right)~. \nonumber\\
&&
\label{eq:MFWrealYY}
\end{eqnarray}

%\subsection{Quasiparticle contribution}

The quasiparticle contribution for the $SU(3)$ case in the mean field approximation reads
\begin{equation}
\Omega_{qp} = 2T\int\frac{d^3k}{(2\pi)^3}~
\log\left\langle\text{det}_A\left(1-L_A~e^{-E/T}\right)\right\rangle~,
\label{eq:SU3qp}
\end{equation}
where the determinant is taken in the adjoint color representation, 
and the adjoint Polyakov line is given by
$L_A = \exp\left[i\left(\theta_3 T_A^3 + \theta_8 T_A^8\right)\right]$,
with $(T_A^b)_{ac}$ the appropriate generators in the adjoint representation. 
We make use of the following equation to express the adjoint loop in terms
of the fundamental one:
\begin{equation}
L_A^{ab} = 2\text{Tr}\left[L_F T_a L_F^\dagger T_b\right]~,
\label{eq:marco1}
\end{equation}
where $T_a$, $T_b$ correspond to the generators of $SU(3)$ in the fundamental representation,
normalized with the condition $\text{Tr}[T_a T_b] = \delta_{ab}/2$.
To compute the functional determinant in Eq.~\eqref{eq:SU3qp}, we chose a representation
in which $L_A$ is diagonal; the eigenvalues of $L_A$ are $\lambda_1 = \lambda_2 =1$,
$\lambda_3 = \lambda_4^* = e^{i(\theta_3 - \theta_8)}$, $\lambda_5 = \lambda_6^* = e^{i(\theta_3 - \theta_9)}$, 
and finally $\lambda_7 = \lambda_8^* = e^{i(\theta_8 - \theta_9)}$, with $\theta_9 = -(\theta_3 + \theta_8)$.

\begin{widetext}
As a result we find
\begin{eqnarray}
\log\left\langle\text{det}_A\left(1-L_A~e^{-E/T}\right)\right\rangle &=& 2\log(1-e^{-E/T}) 
+ \sum_{i=1}^3\log\left(1+e^{-2E/T}-2\left\langle\omega_i \right\rangle e^{-E/T}\right)~,
\label{eq:PAOLO1}
\end{eqnarray}
where we have defined
\begin{eqnarray}
\omega_1 &=& \cos\left(\theta_3 - \theta_8\right)~,\\
\omega_2 &=& \cos\left(\theta_3 + 2\theta_8\right)~,\\
\omega_3 &=& \cos\left(2\theta_3 + \theta_8\right)~,
\label{eq:PAOLO2}
\end{eqnarray}
and the $\langle\dots\rangle$ denotes the average over the gauge group elements as defined as in Eq.~\eqref{eq:MF}. 
\end{widetext}

Our expression for the functional determinant is in agreement with Eq.~(4.5) of~\cite{Zhang:2010kn},
once we put $x=e^{-E/T}$ and change $x\rightarrow -x$ in Eq.~\eqref{eq:SU3qp} to take into account
that in~\cite{Zhang:2010kn}, fermions in the adjoint representation are considered instead of
bosons. It can be proved analitically that 
\begin{equation}
\langle\omega_1\rangle = \langle\omega_2\rangle = \langle\omega_3\rangle = -\frac{1}{3}~,~~~\alpha=0~;
\label{eq:a0}
\end{equation}
moreover, numerically we have checked that
\begin{equation}
\langle\omega_1\rangle = \langle\omega_2\rangle = \langle\omega_3\rangle \approx 1~,~~~\alpha\rightarrow\infty~.
\label{eq:aIN}
\end{equation}

For the sake of reference, we write down the equation for the traced 
fundamental and adjoint Polyakov loops in terms
of the angles $\theta_3$ and $\theta_8$, namely
\begin{eqnarray}
\ell_F &=& \frac{1+\omega_1 + \omega_2 + \omega_3}{4}~, \\
\ell_A &=& \cos\frac{\theta_3-\theta_8}{2}\cos\frac{\theta_3+2\theta_8}{2}\cos\frac{2\theta_3+\theta_8}{2}~.
\label{eq:ADJsu3}
\end{eqnarray}
We have checked numerically that $\langle\ell_F\rangle, \langle\ell_A\rangle \approx 1$ in the asymptotic limit $\alpha\rightarrow\infty$
and $\langle\ell_F\rangle, \langle\ell_A\rangle \rightarrow 0$ for $\alpha\rightarrow 0$ (the latter step can be proved analitically as well).
The asymptotic behavior of the averages of the $\omega_i$ in Eq.~\eqref{eq:aIN} implies that in the deconfinement phase,
at very large temperature where $\langle\ell_A\rangle \approx 1$, the thermal distribution of the quasigluons
approaches that of a perfect gas of massive particles. 
However the thermodynamics of the system remains different from the one of a mere massive gas because of the
Polyakov background mean field.

\section{Numerical Results}
\begin{figure}[t!]
\includegraphics[width=7.5cm]{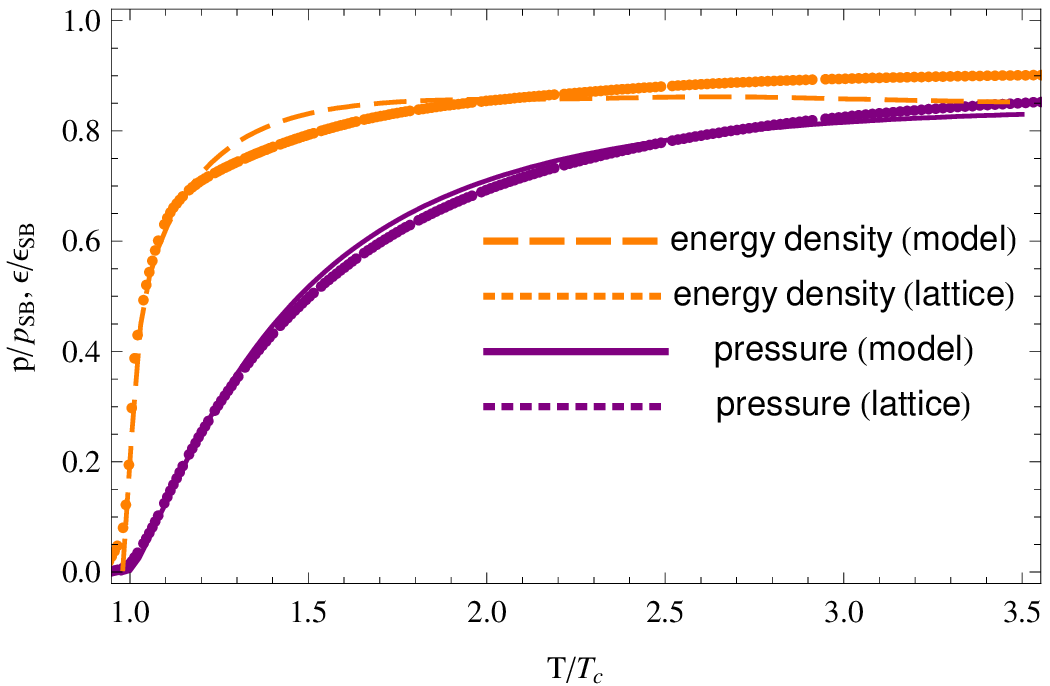}\\
\includegraphics[width=7.5cm]{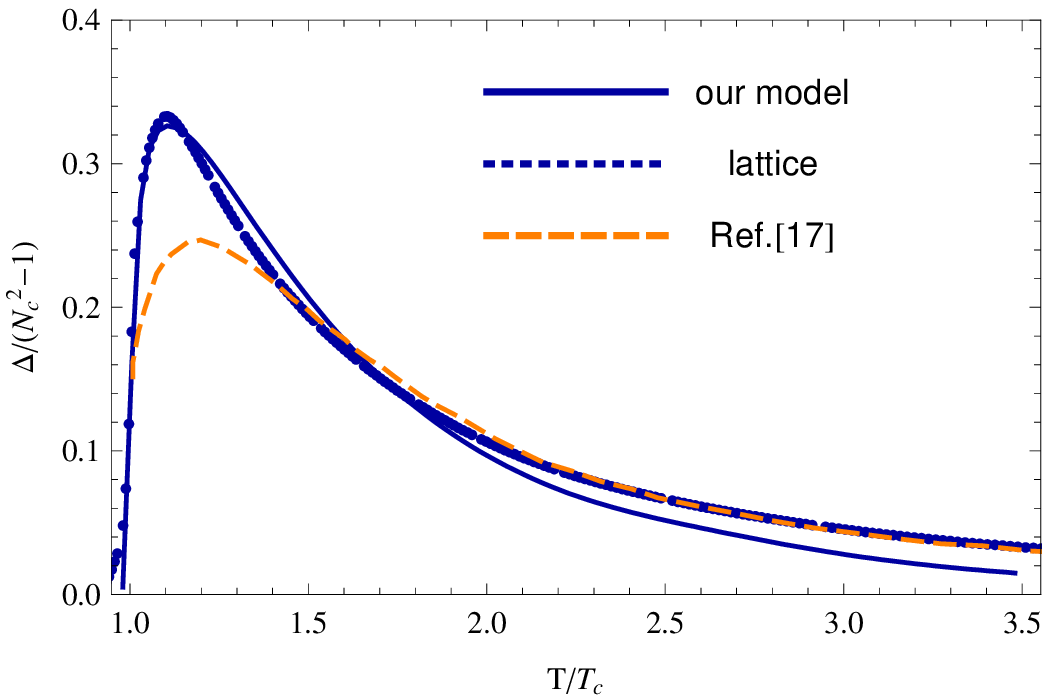}
\caption{\label{Fig:compSU3} {\em Upper panel.} Pressure (indigo, solid line) and energy density (orange, dashed line), 
normalized to the 
respective Stefan-Boltzmann values, as a function of temperature (measured in units of the critical temperature $T_c = 270$ MeV),
for the SU(3) case. Dots correspond to lattice data taken from Ref.~\cite{Boyd:1996bx};
solid lines are the result of our numerical computation within the quasiparticle model.
{\em Lower panel.} Interaction measure per degree of freedon as a function of temperature.
The dashed orange line corresponds to the result of Ref.~\cite{Meisinger:2003id}.}
\end{figure}

In Fig.~\ref{Fig:compSU3} we plot the pressure $p=-\Omega$ (red data) and energy density (green data), normalized to the 
Stefan-Boltzmann value, as a function of temperature (measured in units of the critical temperature),
for the SU(3) case. Given the pressure, the energy density is computed
by virtue of the thermodynamical relation performing the pertinent total derivative of the temperature, $\varepsilon = -p + T dp/dT$.
Of the four parameters in our model, one of them is fixed in order to reproduce the
first order phase transition at $T=270$ MeV; the remaining three parameters are fixed 
in order to require that the mean quadratic deviation for pressure, energy density and interaction
measure between our computation and the Lattice data is minimized.
This procedure leads to the numerical values $a=901.9$ MeV, $b=(157.5$ MeV$)^3$, $w=240$ MeV
and finally $\lambda = 0.25$.  These parameters produce a gluon mass $M=360 \,$MeV at $T = T_c$. 
In the figure, $p_{SB}=d_A \pi^2 T^4/45$ and $\epsilon_{SB}=3 p_{SB}$
where $d_A = N_c^2 -1$ is the dimension of the adjoint representation.
Dots correspond to lattice data taken from Ref.~\cite{Boyd:1996bx};
solid lines are the result of our numerical computation within the quasiparticle model.

In the lower panel of Fig.~\ref{Fig:compSU3} we compare our results for the interaction
measure, $\Delta = (\varepsilon - 3p)/T^4$, with the lattice data (represented by
dots in the figure). We notice that our model reproduces fairly well the peak of the
interaction measure observed on the lattice in the critical region. We also compare
our results with the ones obtained in~\cite{Meisinger:2003id}, where 
a phenomenological potential for the Polyakov loop is considered instead of our
matrix model, and the usual mean field approximation is used. 
The comparison shows that 
%even if the tail is not well reproduced in our case
%(because the matrix model potential gives a nonvanishing contribution
%in the range of temperature considered in this work), 
the peak is much better reproduced in our case. However one could argue that the tail
at high $T$ in instead not so well reproduced; in the next section we will see that such
a drawback can be cured by a slightly different $M_g(T)$ behavior.

\begin{figure}[t!]
\includegraphics[width=7.5cm]{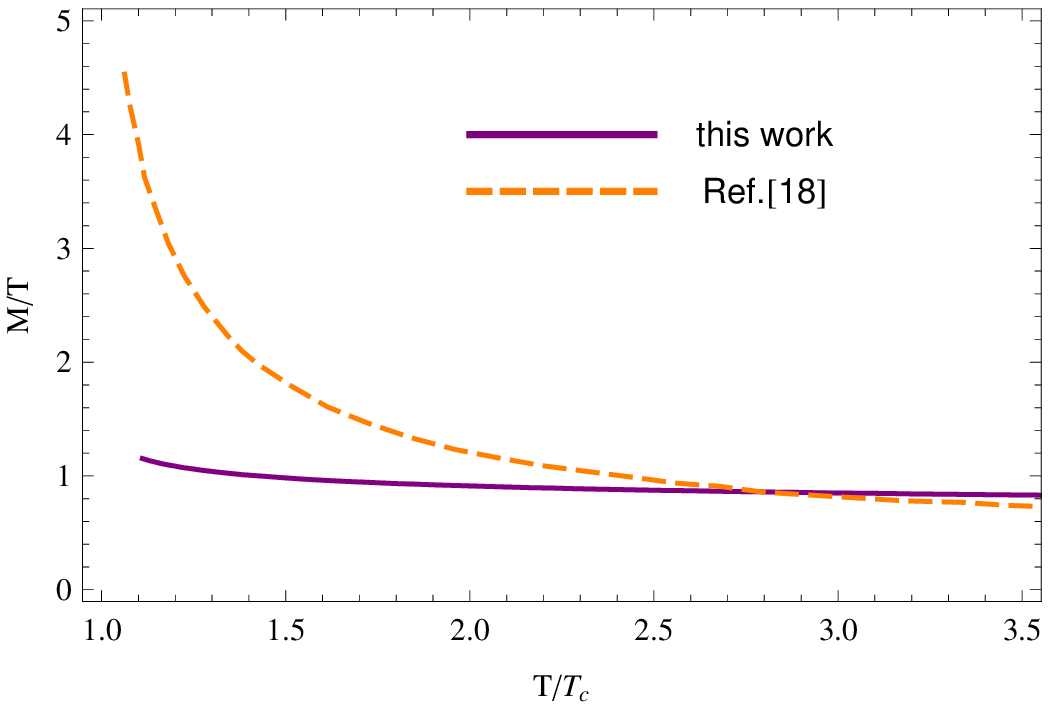}\\
\includegraphics[width=7.5cm]{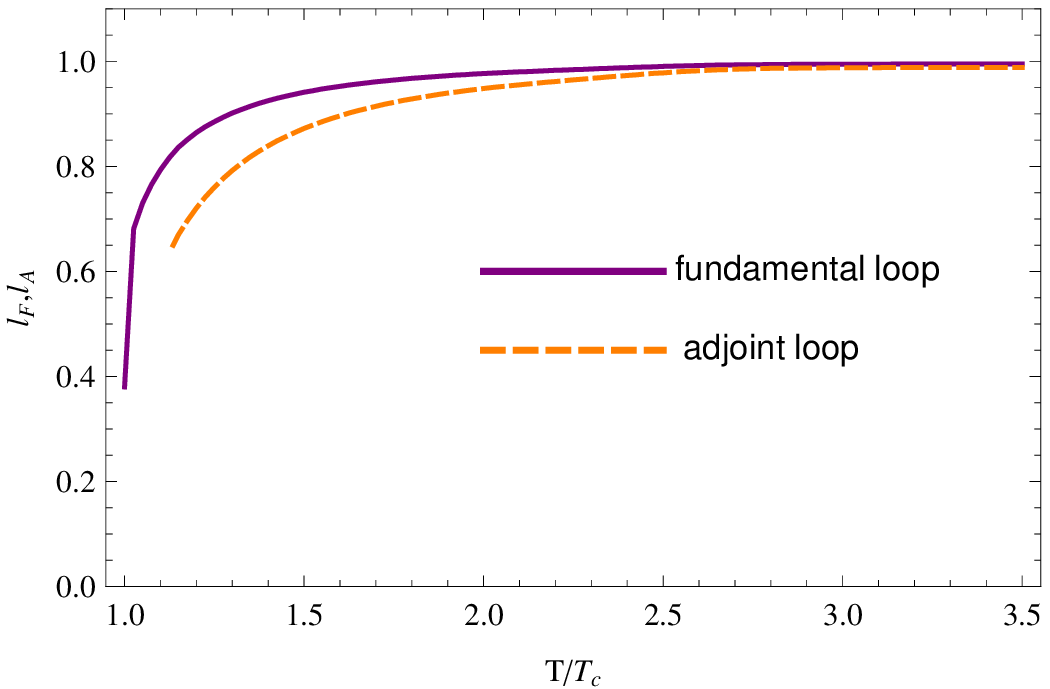}
\caption{\label{Fig:masses} {\em Upper panel}. Ratio $M/T$ for our model (indigo, solid line) and 
for the model without Polyakov loop (orange, dashed line) of~\cite{Castorina:2011ra}, against
temperature, the latter measured in units of the critical temperature $T=270$ MeV.
{\em Lower panel}. Expectation value of the fundamental (indigo, solid line) and adjoint
(orange, dashed line) loops against temperature.}
\end{figure}

In the upper panel of Fig.~\ref{Fig:masses} we plot our result for the gluon mass
against temperature, and we compare it with the result of~\cite{Castorina:2011ra} obtained without the
Polyakov loop dynamics. In Ref.~\cite{Castorina:2011ra} the following ansatz is used:
\begin{equation}
M = \frac{A}{(t-\delta)^c} + B t~,
\label{eq:casto}
\end{equation}
with $t = T/T_c$. Numerical values of the parameters are $A=1.42\, T_c$, $B= 0.533\, T_c$,
$c=0.46$ and $\delta=0.952$. The first addendum on the r.h.s. of the above equation,
which is interpreted as the energy contained in the correlation volume, is important
close to $T_c$, and it is necessary to have a large mass as $T$ approaches $T_c$ from above. 
In fact, when the Polyakov loop
is not introduced, a large mass is needed in order to suppress pressure and energy density around $T_c$. 
In the case the phase transition is of the second order, as it happens for $SU(2)$,
correlation length diverges at $T_c$; therefore, $\delta=1$ since the mass diverges as well. 
When the phase transition is
of the first order, the correlation lenght (hence the correlation volume) 
does not diverge at $T_c$, but it is expected to be large; as a consequence, 
the mass should increase as $T$ approaches $T_c$ from above. For this reason
$\delta$ in Eq.~\eqref{eq:casto} is slightly less than unity $\delta < 1$ . The second addendum on the r.h.s.
of Eq.~\eqref{eq:casto} dominates the mass at large temperature, and its origin
is of perturbative nature.

\begin{figure}[t!]
\includegraphics[width=7.5cm]{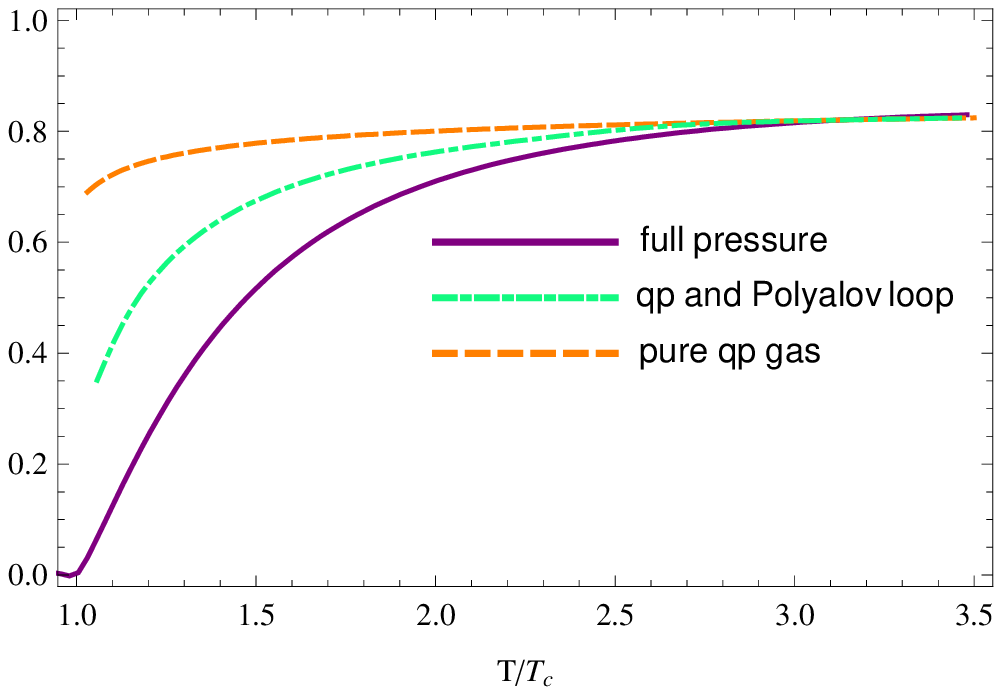}\\
\includegraphics[width=7.5cm]{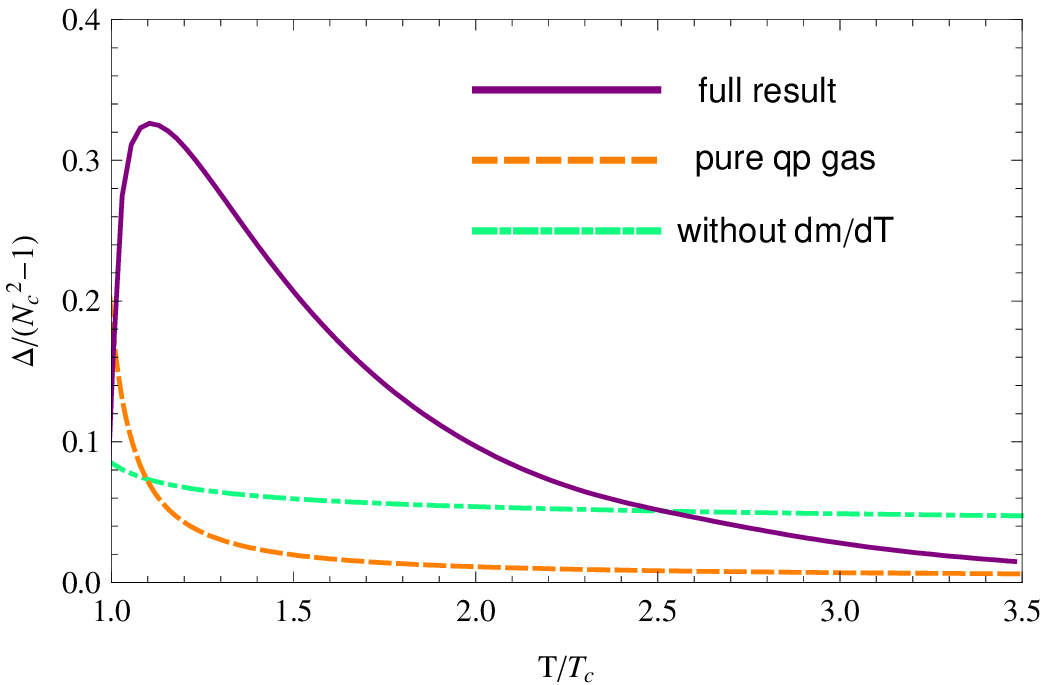}
\caption{\label{Fig:yulia} {\em Upper panel.} Full pressure of our model (indigo, solid line), 
perfect gas pressure (orange, dashed line) and quasiparticle contribution (teal, dot dashed line)
against temperature.
All the data are measured in units
of the Stefan-Boltzmann pressure. {\em Lower panel.} Interaction measure of our model (solid, indigo line) compared with
the pure quasiparticle model (orange, dashed line). In the figure,
the dot dashed teal line corresponds to
the interaction measure we would obtain neglecting the derivative of the gluon mass, see the
text for more detail.}
\end{figure}
 
In our case, the statistical suppression of states below and around $T_c$ is achieved by virtue of the Polyakov loop. 
For this reason, we do not need to have a large mass as $T$ approaches $T_c$.
This mechanism is similar to the statistical confinement that takes place in the
PNJL model~\cite{Fukushima:2008wg,Abuki:2008nm}. To show this aspect
of the model clearly, we have plotted in Fig.~\ref{Fig:yulia} the full
pressure obtained within our model (indigo solid line), and we have compared
this result with the pressure of a perfect gas having the temperature dependent
mass we have obtained in our computations (orange dashed line). From the data shown in the 
figure, the role of the Polyakov loop to suppress thermodynamically the 
contribution of states is noticeable. For completeness, we have also plotted
the quasiparticle gas contribution to the pressure, $\Omega_{qp}$, with Polyakov loop
taken into account (the latter has been selfconsistently computed at any
temperature).

The role of the Polyakov loop is apparent also in the lower panel
of Fig.~\ref{Fig:yulia}, where we compare our results for the interaction measure,
which has Polyakov loop thermodynamics built into it, and the interaction measure for
a gas of quasiparticle with mass given by Eq.~\eqref{eq:mass}. For a matter of
simplicity, the latter is estimated
by using the relativistic Boltzmann distribution function for gluons, namely~\cite{Castorina:2011ja} 
\begin{equation}
\Delta_{B} = 2\frac{(N_c^2 - 1)M^2}{2\pi^2 T^2} 
\left[\frac{M}{T} - \left(\frac{dM}{dT}\right)\right]
K_1(M/T)~,
\label{eq:BOLX}
\end{equation}
%where $K_1$ is a modified second kind Bessel function.
the analytical expression allows to identify the two contributions to
$\Delta_B$, we have verified that turning to the Bose-Einstein statistics does not lead
to significant different results.

It is evident that the Polyakov loop
gives a relevant contribution to the interaction measure. Moreover, it is instructive
to show the concrete effect of a temperature dependent mass on the interaction
measure. To this end, in the figure we have plotted Eq.~\eqref{eq:BOLX} in which
we neglect the term proportional to $dM/dT$: the result is shown by the 
teal line. Since $dM/dT$ in our case is positive, its contribution suppresses
the interaction measure, as it is clear from the data shown in the figure.

\begin{figure}[t!]
\includegraphics[width=7.5cm]{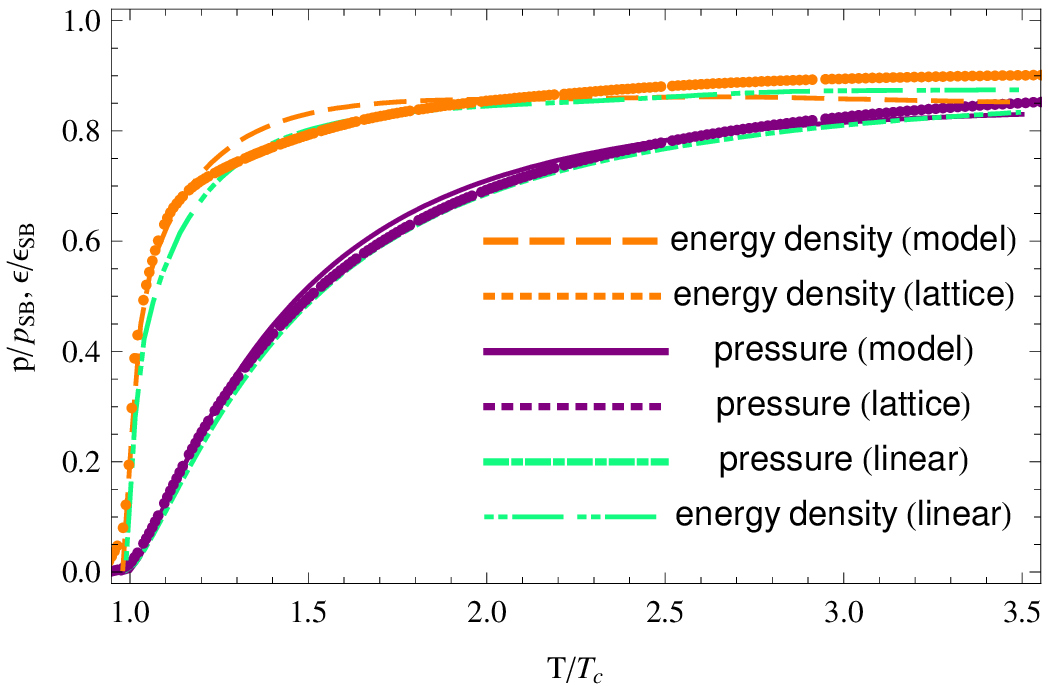}\\
\includegraphics[width=7.5cm]{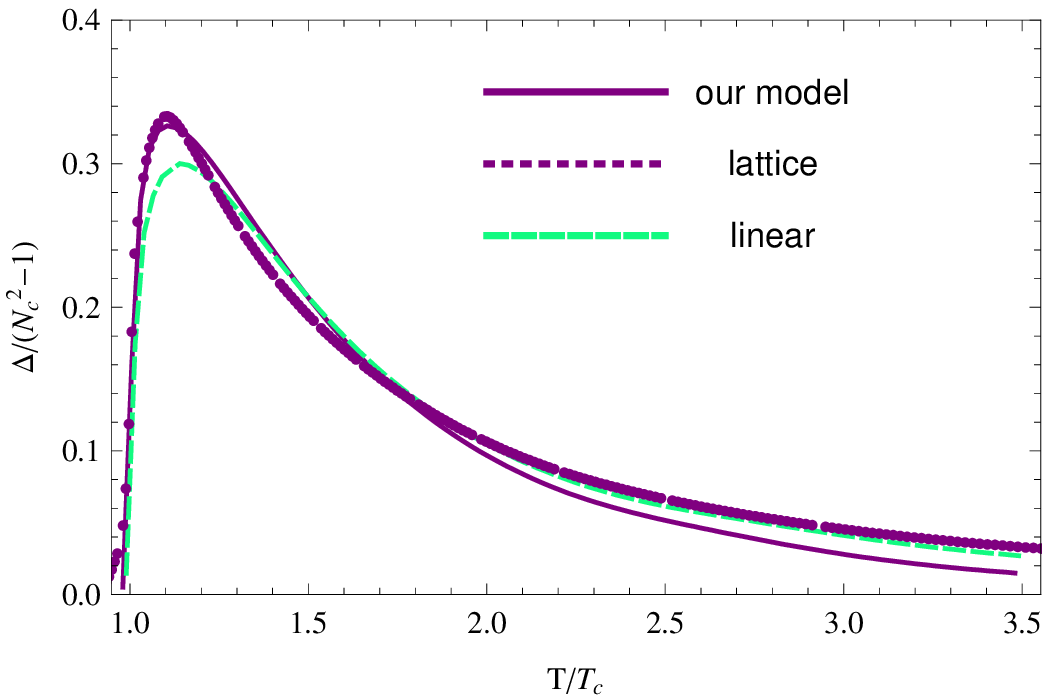}\\
\includegraphics[width=7.5cm]{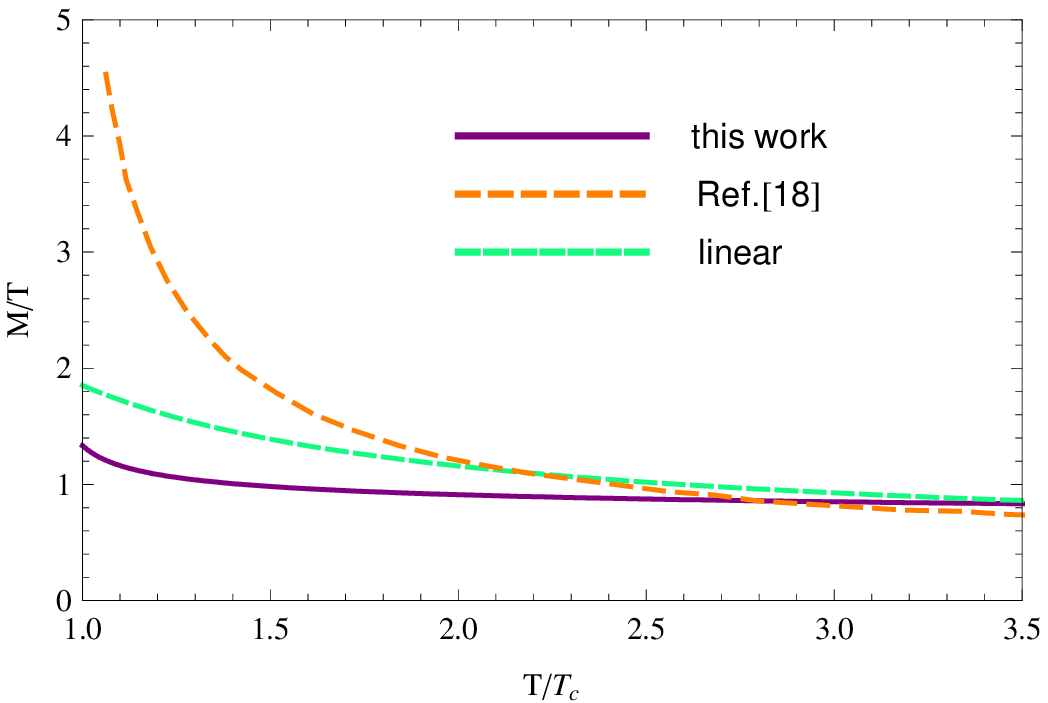}
\caption{\label{Fig:compSU3lin} {\em Upper panel.} Normalized pressure and energy density 
against temperature. 
{\em Middle panel.} Interaction measure per degree of freedon as a function of temperature.
{\em Lower panel.}   The ratio $M/T$ against temperature.}
\end{figure}

As a final investigation, we have considered another ansatz for the thermal
gluon mass, that is
\begin{equation}
M_{linear} = M_0 + h\left(\frac{T}{T_c}-1\right)~,
\label{eq:linearMMM}
\end{equation}
where $M_0 = 500$ MeV and $h=126$ MeV (these two coefficients
have been fixed once again by a minimization procedure of the
quadratic relative deviations). The results are summarized
in Fig.~\ref{Fig:compSU3lin}. Few comments follow.
Secondly, the peak of the interaction
measure is slightly suppressed: in this case the suppression is of about
$10\%$, which has to be compared with the case of the logharithmic mass
(less than $2\%$). However the tail of $\Delta$
is better reproduced, because of the combined effect of mass,
Polyakov loop and $dM/dT$. The role of the Polyakov loop in avoiding the
divergence of the gluon mass as $T_c$ is approached from above remains. 
We leave a more systematic study of 
the several possible ansatzes to a future work.

\section{Conclusions}
In this article, we have studied a combined description of the deconfinement phase of the $SU(3)$ gluon plasma,
in terms of gluon quasiparticles and Polyakov loop. The two are mixed by a phenomenological
potential, see Eq.~\eqref{eq:gqp}, which is inspired by weak coupling computations~\cite{Weiss:1980rj}.
To the quasiparticle contribution, we have added a potential for the Polyakov loop given by Eq.~\eqref{eq:MFWreal};
its analytic form is inspired by the leading order strong coupling expansion, and it has been used in the literature
in the context of the PNJL model~\cite{Abuki:2009dt,Zhang:2010kn}. 

Our main purpose is twofold. Firstly, we are interested to a simple description of
the lattice data about the thermodynamics of the gluon plasma. This is interesting
because it allows to understand which are the relevant degrees of freedom 
in the deconfinement phase of the theory. Moreover, once the proper degrees
of freedom are identified, the effective description studied here can be completed
by adding dynamical quarks.
Employing a relativistic transport theory, this can allow also a direct connection 
between the developments of effective models and the study of the quark-gluon plasma
created in relativistic heavy-ion collisions.

Within our model, we are able to reproduce fairly well the lattice data about
the gluon plasma thermodynamics in the critical region. In particular, we reproduce
the peak of the interaction measure within a 2\%, see lower panel in Fig.~\ref{Fig:compSU3}.
%This result is an improvement of the previous ones of~\cite{Meisinger:2003id}.
%In comparison with~\cite{Meisinger:2003id}, we use a different potential for the Polyakov loop
%and a different mean field procedure. Both of these ingredients are inspired by similar studies
%with dynamical quarks~\cite{Abuki:2009dt,Zhang:2010kn}.

One of the main conclusions of our work is that because of the coupling to the Polyakov loop,
a gluon mass of the order of the critical temperature, $m_g\sim 1-2\, T_c$, is enough to reproduce
the lattice data in the critical region. This is different from what is found
in standard quasi-particle models. The Polyakov loop
is neglected and the gluon plasma is described in terms of a perfect gas of
massive gluons. As a consequence, a mass rapidly increasing with $T$
is needed to suppress pressure and energy density in the critical region. 
On the other hand, in the case studied in this article, the suppression of
the pressure and energy density is mainly caused by the Polyakov loop,
as shown in Fig.~\ref{Fig:yulia}. Therefore, lighter quasiparticles with a smooth 
$T-$dependence
describe fairly well the lattice data in the critical region.

There are several interesting directions to follow to extend the present
work. Firstly, having in mind the study of the quark-gluon plasma,
we plan to add dynamical massive quarks to the picture. 
It is also of a certain interest to span the parameter
space in more detail, eventually analyzing several
functional forms of the gluon mass. 
Moreover, even in the case of the pure gauge theory, it would be interesting
to investigate the way to adjust the interaction
measure in order to reproduce lattice data in the regime of
very large temperature. Even more, it would be of a certain interest
to extend our study to the case of different gauge groups.

{\bf Note added.} While preparing the present article, 
Ref.~\cite{Sasaki:2012bi} was submitted to arXiv, which has 
some overlap with our work.

%\acknowledgments 
%This work was supported in part by the Italian Ministry
%of Education, Universities and Research under the Firb Research Grant
%RBFR0814TT.

\end{document}